# PERFORMANCE ORIENTED QUERY PROCESSING IN GEO BASED LOCATION SEARCH ENGINES

[1] Dr.M.Umamaheswari
Dept. of Computer Science, Bharath University
Chennai-73, Tamil Nadu, India
druma_cs@yahoo.com

[2] S.Sivasubramanian
Dept. of Computer Science, Bharath University
Chennai-73, Tamil Nadu, India
sivamdu2001@Yahoo.Com

*ABSTRACT*

*Geographic location search engines allow users to constrain and order search results in an intuitive manner by focusing a query on a particular geographic region. Geographic search technology, also called location search, has recently received significant interest from major search engine companies. Academic research in this area has focused primarily on techniques for extracting geographic knowledge from the web. In this paper, we study the problem of efficient query processing in scalable geographic search engines. Query processing is a major bottleneck in standard web search engines, and the main reason for the thousands of machines used by the major engines. Geographic search engine query processing is different in that it requires a combination of text and spatial data processing techniques. We propose several algorithms for efficient query processing in geographic search engines, integrate them into an existing web search query processor, and evaluate them on large sets of real data and query traces.*

**I.INTRODUCTION**   The World-Wide Web has reached a size where it is becoming increasingly challenging to satisfy certain information needs. While search engines are still able to index a reasonable subset of the (surface) web, the pages a user is really looking for are often buried under hundreds of thousands of less interesting results. Thus, search engine users are in danger of drowning in information. Adding additional terms to standard keyword searches often fails to narrow down results in the desired direction. A natural approach is to add advanced features that allow users to express other constraints or preferences in an intuitive manner, resulting in the desired documents to be returned among the first results. In fact, search engines have added a variety of such features, often under a special "advanced search" interface, but mostly limited to fairly simple conditions on domain, link structure, or modification date. In this paper we focus on geographic web search engines, which allow users to constrain web queries to certain geographic areas. In many cases, users are interested in information with geographic constraints, such as local businesses, locally relevant news items, or Permission to make digital or hard copies of all or part of this work for personal or classroom use is granted without fee provided that copies are not made or distributed for profit or commercial advantage and that copies bear this notice and the full citation on the first page. To copy otherwise, tore publish, to post on servers or to redistribute to lists, requires prior specific tourism information about a particular region. For example, when searching for yoga classes, local yoga schools are of much higher interest than the web sites of the world's largest yoga schools. We expect that '*geographic search engine's*, that is, search engines that support geographic preferences, will have a major impact on search technology and their business models. First, geographic search engines provide a very useful tool. They allow users to express in a single query what might take multiple queries with a standard search engine.

A. LOCATION BASED

A user of a standard search engine looking for a yoga school in or close to Tambaram, Chennai, might have to try queries such as

- yoga ''Delhi''
- yoga "Chennai"
- yoga''Tambaram'' (a part of Chennai)

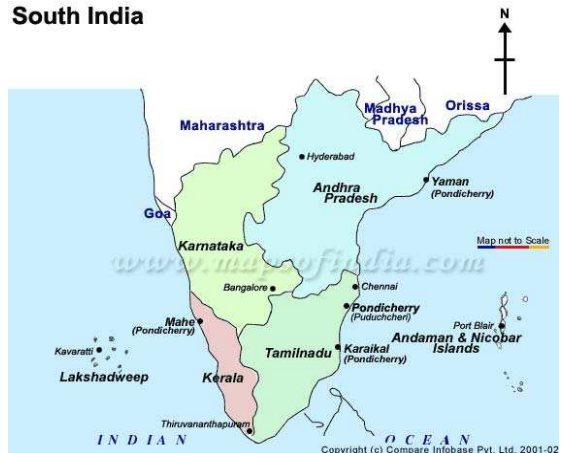

but this might yield inferior results as there are many ways to refer to a particular area, and since a purely text-based engine has no notion of geographical closeness (example, a result across the bridge to Tambaram or nearby in Gundy might also be acceptable). Second, geographic search is a fundamental enabling technology for "*location-based services*", including electronic commerce via cellular phones and other mobile devices. Third, geographic search supports locally targeted web advertising, thus attracting advertisement budgets of small businesses with a local focus. Other opportunities arise from mining geographic properties of the web, example, for market research and competitive intelligence. Given these opportunities, it comes as no surprise that over the last two years leading search engine companies such as *Google* and *Yahoo* have made significant efforts to deploy their own versions of geographic web search. There has also been some work by the academic research community, to mainly on the problem of extracting geographic knowledge from web pages and queries. Our approach here is based on a setup for geographic query processing that we recently introduced in [1] in the context of a geographic search engine prototype. While there are many different ways to formalize the query processing problem in geographic search engines, we believe that our approach results in a very general framework that can capture many scenarios.

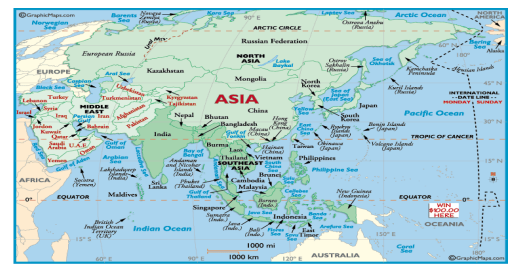

B. QUERY FOOTPRINT

We focus on the efficiency of query processing in geographic search engines, example, how to maximize the query throughput for a given problem size and amount of hardware. Query processing is the major performance bottleneck in current standard web search engines, and the main reason behind the thousands of machines used by larger commercial players. Adding geographic constraints to search queries results in additional challenges during query execution which we now briefly outline. In a nutshell, given a user query consisting of several keywords, a standard search engine ranks the pages in its collection in terms of their relevance to the keywords. This is done by using a text index structure called an inverted index to retrieve the IDs of pages containing the keywords, and then evaluating a term-based ranking function on these pages to determine the $k$ highest-scoring pages. (Other factors such as hyperlink structure and user behavior are also often used, as discussed later). Query processing is highly optimized to exploit the properties of inverted index structures, stored in an optimized compressed format, fetched from disk using efficient scan operations, and cached in main memory. In contrast, a query to a geographic search engine consists of keywords and the geographic area that interests the user, called "*query footprint*".

Each page in the search engine also has a geographic area of relevance associated with it, called the '*geographic footprin't* of the page. This area of relevance can be obtained by analyzing the collection in a preprocessing step that extracts geographic information, such as city names, addresses, or references to landmarks, from the pages and then maps these to positions using external geographic databases. In other approaches it is assumed that this information is provided via meta tags or by third parties. The resulting page footprint is an arbitrary, possibly noncontiguous area, with an amplitude value specifying the degree of relevance of each location. Footprints can be represented as polygons or bitmap-based structures; details of the representation are not important here. A geo search engine computes and orders results based on two factors.

**C. KEYWORDS AND GEOGRAPHY.**

Given a query, it identifies pages that contain the keywords and whose page footprint intersects with the query footprint, and ranks these results according to a combination of a term-based ranking function and a geographic ranking function that might, example, depend on the volume of the intersection between page and query footprint. Page footprints could of course be indexed via standard spatial indexes such as $R*$-trees, but how can such index structures be integrated into a search engine query processor, which is optimized towards inverted index structures? How should the various structures be laid out on disk for maximal throughput, and how should the data flow during query execution in such a mixed engine? Should we first execute the textual part of the query, or first the spatial part, or choose a different ordering for each query? These are the basic types of problems that we address in this paper. We first provide some background on web search engines and geographic web search technology. We assume that readers are somewhat familiar with basic spatial data structures and processing, but may have less background about search engines and their inner workings. Our own perspective is more search-engine centric: given a high-performance search engine query processor developed in our group, our goal is to efficiently integrate the types of spatial operations arising in geographic search engines

## II. BASICS OF SEARCH ENGINE ARCHITECTURE

The basic functions of a crawl-based web search engine can be divided into 'crawling, data mining, index construction, and query processing'. During crawling, a set of initial seed pages is fetched from the web, parsed for hyperlinks, and then the pages pointed to by these hyperlinks are fetched and parsed, and so on, until a sufficient number of pages has been acquired. Second, various data mining operations are performed on the acquired data, example, detection of web spam and duplicates, link analysis based on Page rank [7], or mining of word associations. Third, a text index structure is built on the collection to support efficient query processing. Finally, when users issue queries, the top-10 results are retrieved by traversing this index structure and ranking encountered pages according to various measures of relevance. Search engines typically use a text index structure called an *inverted index*, which allows efficient retrieval of documents containing a particular word (*term*). Such an index consists of many *inverted lists*, where each inverted list $I_w$ contains the IDs of all documents in the collection that contain a particular word $w$, usually sorted by document ID, plus additional information about each occurrence. Given, example, a query containing the search terms "apple"," orange", and "pear", a search engine traverses the inverted list of each term and uses the information embedded therein, such as the number of search term occurrences and their positions and contexts, to compute a score for each document containing the search terms. We now formally introduce some of these concepts.

### A. DOCUMENTS, TERMS, AND QUERIES:

We assume a collection $D = \{d0, d1, \ldots dn-1\}$ of $n$ web pages that have been crawled and are stored on disk. Let $W = \{w0, w1, \ldots, wm-1\}$ be all the different words that occur anywhere in *D*. Typically, almost any text string that appears between separating symbols such as spaces, commas, etc., is treated as a valid word (or *term*). A query

$$q = \{t0, t1, \ldots, td-1\} \qquad (1)$$

is a set1 of words (terms).

### B. INVERTED INDEX:

An inverted index I for the collection consists of a set of inverted lists

$$Iw0, Iw1, \ldots, Iwm-1 \qquad (2)$$

Where list *Iw* contains a *posting* for each occurrence of word *w*. Each posting contains the ID of the document where the word occurs, the position within the document, and possibly some context (in a title, in large or bold font, in an anchor text). The postings in each inverted list are usually sorted by document IDs and laid out sequentially on disk, enabling efficient retrieval and decompression of the list. Thus, Boolean queries can be implemented as unions and intersections of these lists, while phrase searches

### C. TERM-BASED RANKING:

The most common way to perform ranking is based on comparing the words (terms) contained in the document and in the query. More precisely, documents are modeled as unordered bags of words, and a ranking function assigns a score to each document with respect to the current query, based on the frequency of each query word in the page and in the overall collection, the length of the document, and maybe the context of the occurrence (example, higher score if term in title or bold face). Formally, given "a query (1) is", a *ranking function F* assigns to each document *D* a score *F (D, q)*. The system then returns the *k* documents with the highest score. One popular class of ranking functions is the *cosine measure* [44], for example

$$F(D,q) = \sum_{i=0}^{d-1} \frac{\ln(1+n/f_{t_i}) \cdot 1 + \ln f_{D,t_i}}{\sqrt{|D|}}, \quad (3)$$

In the equation (3) Where *fD,ti* and *fti* are the frequency of term *ti* in document *D* and in the entire collection, respectively. Many other functions have been proposed, and the techniques in this paper are not limited to any particular class. In addition, scores based on link analysis or user feedback are often added into the total score of a document; in most cases this does not affect the overall query execution strategy if these contributions can be pre computed offline and stored in a memory-based table or embedded into the index. For example, the ranking function might become something like *F (D, q) = pr(D)+F(D, q)* where *pr(D)* is a pre computed and suitably normalized Page rank score of page *D*. The key point is that the above types of ranking functions can be computed by first scanning the inverted lists associated with the search terms to find the documents in their intersection, and then evaluating the ranking function only on those documents, using the information embedded in the index. Thus, at least in its basic form, query processing with inverted lists can be performed using only a few highly efficient scan operations, without any random lookups.

### III. BASICS OF GEOGRAPHIC WEB SEARCH

We now discuss the additional issues that arise in a geographic web search engine. Most details of the existing commercial systems are proprietary; our discussion here draws from the published descriptions of academic efforts in [1, 3] the first task, crawling, stays the same if the engine aims to cover the entire web. In our systems we focus on Germany and crawl the de domain; in cases where the coverage area does not correspond well to any set of domains, focused crawling strategies [4 may be needed to find the relevant pages.

**A. GEO CODING:** Additional steps are performed as part of the data mining task in geographical search engines, in order to extract geographical information from the collection. Recall that the footprint of a page is a potentially noncontiguous area of geographical relevance. For every location in the footprint, an associated integer value expresses the *certainty* with which we believe the page is actually relevant to the location. The process of determining suitable geographic footprints for the pages is called 'geo coding' [3] In [1], geo coding consists of three steps, *geo extraction*, *geo matching*, and *geo propagation*. The first step extracts all elements from a page that indicate a location, such as city names, addresses, landmarks, phone numbers, or company names. The second step maps the extracted elements to actual locations (that is, coordinates), if necessary resolving any remaining ambiguities, example, between cities of the same name. This results in an initial set of footprints for the pages. Note that if a page contains several geographic references, its footprint may consist of several noncontiguous areas, possibly with higher certainty values resulting, say, from a complete address at the top of a page or a town name in the URL than from a single use of a town name somewhere else in the page text.

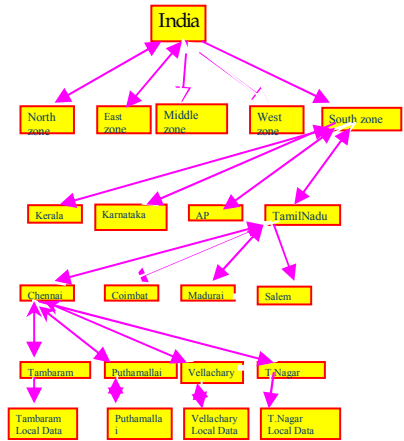

**Figure 1.1 shows an example of a page and its split footprint.**

The third step, *geo propagation*, improves quality and coverage often initial geo coding by analysis of link structure and site topology. Thus, a page on the same site as many pages relevant to Chennai City, or with many hyperlinks to or from such pages, is also more likely to be relevant to Chennai and should inherit such a footprint (though with lower certainty). In addition, geo coding might exploit external data sources such as whois data, yellow pages, or regional web directories. The result of the data mining phase is a set of footprints for the pages in the collection. In [30], footprints were represented as which were stored in a highly compressed quad-tree structure, but this decision is not really of concern to us here. Other reasonably compact and efficient representations, example, as polygons, would also work. All of our algorithms approximate the footprints by sets of bounding rectangles; we only assume the existence of a black-box procedure for computing the precise geographical score between a query footprint and a document footprint. During index construction, additional spatial index structures are created for document footprints as described later.

**B. GEOGRAPHIC QUERY PROCESSING:** As in [1], each search query consists of a set of (textual) terms, and a query footprint that specifies the geographical area of interest to the user. We assume a geographic ranking function that assigns a score to each document footprint with respect to the query footprint, and that is zero if the intersection is empty; natural choices are the inner product or the volume of the intersection. Thus, our overall ranking function might be of the form $F(D, q) = g(fD, fq) + pr(D) + F(D, q)$, with a term-based ranking function $F(D, q)$, a global rank $pr(D)$ (example, Pagerank), and a geographic score $g(fD, fq)$ computed from query footprint $fq$ and document footprint $fD$ (with appropriate normalization of the three terms). Our focus in this paper is on how to efficiently compute such ranking functions using a combination of text and spatial index structures. Note that the query footprint can be supplied by the user in a number of ways. For mobile devices, it seems natural to choose a certain area around the current location of the user as a default footprint. In other cases, a footprint could be determined by analyzing a textual query for geographic terms, or by allowing the user to click on a map. This is an interface issue that is completely orthogonal to our approach.

## IV. ALGORITHMS

**A. TEXT-FIRST BASELINE:** This algorithm first filters results according to textual search terms and thereafter according to geography. Thus, it first accesses the inverted index, as in a standard search engine, retrieving a sorted list of the docIDs (and associated data) of documents that contain all query terms. Next, it retrieves all footprints of these documents. Footprints are arranged on disk sorted by docID, and a reasonable disk access policy is used to fetch them: footprints close to each other are fetched in a single access, while larger gaps between footprints on disk are traversed via a forward seek. Note that in the context of a DAAT text query processor, the various steps in fact overlap. The inverted index access results in a sorted stream of docIDs for documents that contain all query terms, which is directly fed into the retrieval of document footprints, and precise scores are computed as soon as footprints arrive from disk.

**B. GEO-FIRST BASELINE:** This algorithm uses a spatial data structure to decrease the number of footprints fetched from disk. In particular, footprints are approximated by MBRs that (together with their corresponding docIDs) are kept in a small (memory-resident) R∗-tree. As before, the actual footprints are stored on disk, sorted by docID. The algorithm first accesses the R∗-tree to obtain the docIDs of all documents whose footprint is likely to intersect the query footprint. It sorts the docIDs, and then filters them by using the inverted index. Finally, it fetches the remaining footprints from disk, in order to score documents precisely.

## C. K-SWEEP ALGORITHM

The main idea of the first improved algorithm is to retrieve all required toe print data through a fixed number $k$ of contiguous scans from disk. In particular, we build a grid-based spatial structure in memory that contains for each tile in a $1024 \times 1024$ domain a list of $m$ *toe print ID intervals*. For example, for $m = 2$ a tile $T$ might have two intervals $[3476, 3500]$ and $[23400, 31000]$ that indicate that all toe prints that intersect this tile have toe print IDs in the ranges $[3476, 3500]$ and $[23400, 31000]$. In the case of a $1024 \times 1024$ grid, including about 50% empty tiles, the entire auxiliary structure can be stored in a few MB. This could be reduced as needed by compressing the data or choosing slightly larger tiles (without changing the resolution of the actual footprint data). Given a query, the system first fetches the interval information for all tiles intersecting the query footprint, and then computes up to $k \geq m$ larger intervals called *sweeps* that cover the union of the intervals of these tiles. Due to the characteristics of space filling curves, each interval is usually fairly small and intervals of neighboring tiles overlap each other substantially. As a result, the $k$ generated sweeps are much smaller than the total toe print data. The system next fetches all needed toe print data from disk, by means of $k$ highly efficient scans. The IDs of the encountered toe prints are then translated into doc IDs and sorted. Using the sorted list of docIDs, we then access the inverted index to filter out documents containing the textual query terms. Finally we evaluate the geographic score between the query footprint and the remaining documents and their footprints. The algorithm can be summarized as follows:

*K-SWEEP ALGORITHM:*

(1) Retrieve the toe print ID intervals of all tiles intersecting the

Query footprint.

(2) Perform up to *k* sweeps on disk, to fetch all toe prints in the union of intervals from Step (1).

(3) Sort the doc IDs of the toe prints retrieved in Step (2) and access the inverted index to filter these doc IDs.

(4) Compute the geo scores for the remaining doc IDs using the toe prints retrieved in Step (2).

One limitation of this algorithm is that it fetches the complete data of all toe prints that intersect the query footprint (plus other close by toe prints), without first filtering by query terms. Note that this is necessary since our simple spatial data structure does not contain the actual docIDs for toe prints intersecting the tile. Storing a list of docIDs in each tile would significantly increase the size of the structure as most docIDs would appear in multiple tiles. Thus, we have to first access the toe print data on disk to obtain candidate docIDs that can be filtered through the inverted index

**CONCLUSIONS**

We discussed a general framework for ranking search results based on a combination of textual and spatial criteria, and proposed several algorithms for efficiently executing ranked queries on very large collections. We integrated our algorithms into an existing high-performance search engine query processor and evaluated them on a large data set and realistic geographic queries. Our results show that in many cases geographic query processing can be performed at about the same level of efficiency as text-only queries. There are a number of open problems that we plan to address. Moderate improvements in performance should be obtainable by further tuning of our implementation. Beyond these optimizations, we plan to study pruning techniques for geographic search engines that can produce top-*k* results without computing the precise scores of all documents in the result set. Such techniques could combine early termination approaches from search engines with the use of approximate (lossy-compressed) footprint data. Finally, we plan to study parallel geographic query processing on clusters of machines. In this case, it may be preferable to assign documents to participating nodes not at random, as commonly done by standard search engines, but based on an appropriate partitioning of the underlying

| *Searching of Data* | *Draw back of old system* | *Advantage of Proposed system* |
|---|---|---|
| Accuracy of Local data | Very less local data | Accuracy and more local data |
| Processing time | 0.65 seconds | 0.34 seconds |
| Regional specification | --------NIL---- | Splitting different type of Region |
| Spatial data structure | No link between text and spatial data | Good link between text and spatial data |

**REFERENCES**

[1]. A. Markowetz, Y.-Y. Chen, T. Suel, X. Long, and B. Seeger. Design and implementation of a geographic search

engine. In *8th Int. Workshop on the Web and Databases (WebDB)*, June 2005.

[2]. V. Anh, O. Kretser, and A. Moffat.Vector-space ranking with effective early termination. In *Proc. of the 24th Annual SIGIR Conf. on Research and Development in Information Retrieval*, pages 35–42, September 2001.

[3]. K. McCurley. Geospatial mapping and navigation of the web. In *Proc. of the 10th Int. World Wide Web Conference*, pages 221–229,May 2001.

[4]. S. Chakrabarti, M. van den Berg, and B. Dom. Focused crawling: Anew approach to topic-specific web resource discovery. In *Proc. of the 8th Int. World Wide Web Conference*, May 1999.

[5]. Y. Zhou, X. Xie, C. Wang, Y. Gong, and W. Ma. Hybrid index structures for location-based web search. In *Proc. of the 14th Conf.on Information and Knowledge Management (CIKM)*, pages 155–162, November 2005.

[6]. Reference for query processing in web search engine based on the Journal for Yen-Yu Chen Polytechnic University Brooklyn, NY 11201, USA Torsten Suel Polytechnic University Brooklyn, NY 11201, USA June 2006

**AUTHOR'S PROFILE:**

1)

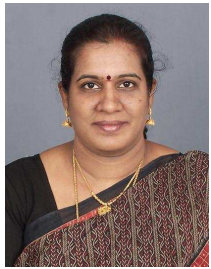

Dr.M.Uma Maheswari received her Bachelor of science in Computer Science from Bharathidasan university in 1995, Master of Computer Applications in Computer Science from Bharathidasan University in1998,M.Phil in Computer Science from Alagappa University, Karaikudi, Master of Technology in Computer Science from Mahatma Gandhi Kasi Vidyapeeth university in 2005 and Ph.D in Computer Science from Magadh Universty, Bodh Gaya in 2007.She has more than 10 years of teaching experience and guided 150 M.C.A projects,23 M.Tech projects and 6 PhD research works.153 -

2)

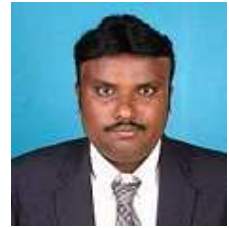

Mr.S.Sivasubramanian received his Diploma in Hardware Software installing in ECIL-BDPS, Govt of India, and Advanced Diploma in computer Application in UCA, Madurai, and Bachelor of Science in Physics from Madurai Kamaraj University in 1995, Master of Science in Physics from Madurai Kamaraj University in 1997, Post Graduate Diploma in Computer and Application in Government of India 2000. Master of Technology in Computer Science and engineering from Bharath University Chennai 2007. Pursing PhD in Bharath University Chennai. He has more than 5 years of teaching experience and guided 20 B.Tech projects, 11 M.Tech projects